\documentclass[a4paper,11pt]{article}

\usepackage{hep-title}
\usepackage[oldstyle]{hep-font}
\usepackage{hep-text}
\usepackage{hep-math-font}
\usepackage{hep-math}
\usepackage{hep-bibliography}
\usepackage{hep-float}
\usepackage[feynman,plot]{hep-graphic}
\usepackage{hep-reference}
\usepackage{hep-acronym}

\usepackage{fullpage}

\usepackage[all]{hypcap}


\usepackage{dcolumn}
\newcolumntype{d}[1]{D{.}{.}{#1}}
\newcommand{\header}[1]{\multicolumn{1}{c}{#1}}

\graphicpath{graphics}
\tikzsetexternalprefix{figures/}

\usepgfplotslibrary{polar}
\usepgfplotslibrary{groupplots}

\pgfplotsset{
    table/search path=data,
    theory/.style={
        mark=none,
        samples=2000,
    },
    data/.style={
        solid,
        thin,
        only marks,
    },
}

\newcommand*{\feynrules}{\software{FeynRules}}
\newcommand*{\feynrulescite}{\cite{Christensen:2008py}}
\newcommand*{\feynrulesver}{\software[2.3]{FeynRules} \feynrulescite}

\newcommand*{\madgraph}{\software{MadGraph5\_aMC@NLO}}
\newcommand*{\madgraphcite}{\cite{Alwall:2014hca}}
\newcommand*{\madgraphver}{\software[3.4.1]{MadGraph5\_aMC@NLO} \madgraphcite}

\newcommand*{\pythiacite}{\cite{Bierlich:2022pfr}}
\newcommand*{\pythiaver}{\software[8.306]{Pythia} \pythiacite}

\newcommand*{\delphes}{\software{Delphes}}
\newcommand*{\delphescite}{\cite{deFavereau:2013fsa}}
\newcommand*{\delphesver}{\software[3.5]{Delphes} \delphescite}

\acronym{SM}{Standard Model}

\acronym{EW}{electroweak}
\acronym{EWSB}{eletroweak symmetry breaking}
\acronym{VEV}{vacuum expectation value}

\acronym{TOF}{time of flight}

\acronym{LO}{leading order}

\acronym{SPSS}{symmetry-protected seesaw scenario}
\acronym{pSPSS}{phenomenological symmetry-protected seesaw scenario}

\acronym{BM}{benchmark model}

\acronym{HNL}{heavy neutral lepton}
\acronym[{}$N\mkern-1.5mu\widebar N$O]{NNO}{heavy neutrino-antineutrino oscillation}
\acronym{FBA}{forward-backward asymmetry}
\acronymalternative{FBA}{fba}{forward-backward asymmetric}

\acronym{GUT}{grand unified theory}
\acronym{LNC}{lepton number conservation}
\acronymalternative{LNC}{lnc}{lepton number conserving}
\acronym{LNV}{lepton number violation}
\acronymalternative{LNV}{lnv}{lepton number violating}
\acronym{LNLS}{\emph{lepton number}-like symmetry}

\newcommand{\qty}[2]{\unit[#1]{#2}}

\newcommand{\num}[1]{\ensuremath{#1}}

\newcommand*{\hc}{\text{H.c.}}

\acronym[FCC-$ee$]{FCCee}{future circular $e^+e^-$ collider}
\acronym{IDEA}{innovative detector for electron-positron accelerator}

\title{Heavy neutrino-antineutrino oscillations at the FCC-$ee$}

\author[basel]{Stefan Antusch}
\author[CFTP]{Jan Hajer}
\author[lisboa]{Bruno M. S. Oliveira}

\affiliation[basel]{Departement Physik, Universität Basel, Klingelbergstrasse 82, CH-4056 Basel, Switzerland}
\affiliation[CFTP]{Centro de Física Teórica de Partículas (CFTP), Instituto Superior Técnico (IST), Universidade de Lisboa, 1049-001 Lisboa, Portugal}
\affiliation[lisboa]{Departamento de Física, Instituto Superior Técnico (IST), Universidade de Lisboa, 1049-001 Lisboa, Portugal}

\begin{document}

\maketitle

\begin{abstract}
    We discuss the impact of \NNOs on \HNL searches at proposed electron-positron colliders such as the \FCCee.
    During the $Z$ pole run, \HNLs can be produced alongside a light neutrino or antineutrino that escapes detection and can decay into a charged lepton or antilepton together with an off-shell $W$ boson.
    In this case, signals of \LNV only show up in the final state distributions.
    We discuss how \NNOs, a typical feature of collider-testable low-scale seesaw models where the heavy neutrinos form pseudo-Dirac pairs, modify such final state distributions.
    For example, the \FBA of the reconstructed heavy (anti)neutrinos develops an oscillatory dependence on the \HNL lifetime.
    We show that these oscillations can be resolvable for long-lived \HNLs.
    We also discuss that when the \NNOs are not resolvable, they can nevertheless significantly modify the theory predictions for \FBAs and observables such as
    the ratio of the total number of \HNL decays into $\ell^-$ over ones into $\ell^+$, in an interval of the angle~$\theta$ between the \HNL and the beam axis.
    Our results show that \NNOs should be included in collider simulations of \HNLs at the \FCCee.
\end{abstract}

\tableofcontents

\section{Introduction}

The observation of neutrino flavour oscillations can be explained by introducing at least two heavy right-chiral neutrinos that generate tiny neutrino masses \cite{SNO:2002tuh}.
Since these additional neutrinos are singlets under the symmetries of the \SM, they can form a Majorana mass term \cite{Majorana:1937vz} in addition to the Dirac mass term involving the \SM Higgs and lepton doublets.
The type~I seesaw mechanism exploits the interplay of these two types of mass terms to explain the smallness of the observed light neutrino masses \cite{Minkowski:1977sc}.

When the measured light neutrino mass scale is generated by a single additional right-chiral neutrino, two limiting cases arise.
On the one hand, if the additional neutrino's Yukawa coupling is of order one, its Majorana mass is pushed close to the \GUT scale.
On the other hand, a Majorana mass around the \EW scale is achieved at the cost of tiny Yukawa couplings that render observation at colliders very challenging.
\footnote{
    At the (HL-)LHC, this \emph{small coupling limit} of the low-scale seesaw mechanism is not testable.
    During the $Z$~pole run of the $\FCCee$, a few events could be observed within a very limited mass range.
        However, the possible event number is too low for a detailed study of the final-state distributions of the $\HNL$ decay products.
}

When adding instead the two right-chiral neutrinos required to explain the observed light neutrino mass squared differences, a third possibility emerges.
It employs a protective \LNLS to ensure that the two \HNLs form an almost degenerate pseudo-Dirac pair; see \cite{Antusch:2022ceb} for a more detailed discussion.
This construction ties the size of the neutrino mass scale to the amount of symmetry breaking.
It allows the Yukawa coupling and the Majorana mass to be in the range where \HNLs are discoverable at collider experiments.

The present work focuses on such collider-testable \SPSSs, namely the recently introduced \pSPSS \cite{Antusch:2022ceb, FR:pSPSS}.
This benchmark scenario introduces two additional parameters over the symmetric limit of the \SPSS \cite{Antusch:2015mia, Antusch:2016ejd}, the heavy neutrino \emph{mass splitting} $\Delta m$ and the effective \emph{damping parameter} $\lambda$.
With these, it captures the collider phenomenology of a pseudo-Dirac pair of \HNLs, namely the \NNOs (\cite{Antusch:2020pnn, Antusch:2022ceb}, see also \cite{Cvetic:2015ura, Anamiati:2016uxp, Antusch:2017ebe}) which lead to the appearance of \LNV.
When $\Delta m$ is much smaller than the average \HNL mass $m$, only a small degeneracy between the two heavy neutrino mass eigenstates is introduced, which causes their interference as they propagate between production and decay.
This interference causes oscillations between the heavy neutrino and antineutrino interaction eigenstates produced via $W$ or $Z$ bosons together with a \SM antilepton and lepton, respectively.
When the heavy neutrinos and antineutrinos finally decay into \SM leptons and antileptons, respectively, the \NNOs give rise to an oscillating pattern of \lnc and \lnv processes as a function of the \HNL lifetime.
\footnote{
    The possibility of resolving the oscillations at the HL-LHC has been discussed in \cite{Antusch:2022hhh, Antusch:2017ebe}.
}
The parameter $\lambda$ accounts for possible decoherence effects, \cf \cite{Antusch:2023nqd}.

The $Z$ pole run of the \FCCee is one of the most promising paths in the search for \HNLs \cite{FCC:2018evy} due to its sensitivity to displaced vertices from \HNLs with masses below the $W$ mass.
Nevertheless, since \HNLs are produced with a light (anti)neutrino that escapes detection, the effects of \LNV are generally difficult to probe and require measuring final state distributions.

One observable that has been discussed in this context for the \BMs of a pure Dirac pair of \HNLs and a single Majorana \HNL is a \FBA \cite{Blondel:2021mss}.
It arises because the $Z$ bosons produced from colliding ultrarelativistic electrons and positrons have a polarisation $P_Z$ of order \unit[15]{\%}.
For Dirac \HNLs, this, in turn, affects the direction in which neutrinos and antineutrinos are emitted.
For pseudo-Dirac \HNLs, \NNOs induce oscillating signatures, significantly modifying the relevant observables.

This paper is organised as follows:
In \cref{sec:pSPSS}, we briefly review the \pSPSS benchmark scenario.
Examples of modified observables in the presence of \NNOs are discussed in \cref{sec:observables}.
Details about our simulations are provided in \cref{sec:simulations}.
In \cref{sec:results}, we present our results for selected \BM parameter points and \cref{sec:conclusions} contains the conclusions.

\section{The $\pSPSS$ benchmark scenario}\label{sec:pSPSS}

\resetacronym{pSPSS}

The \pSPSS \cite{Antusch:2022ceb} is designed to describe the \LO collider phenomenology, including the \NNOs, of the dominant pseudo-Dirac pair of \HNLs with the minimal number of parameters.
As discussed above, pseudo-Dirac \HNLs generically appear in collider-testable low-scale seesaw models where an approximate \LNLS ensures the smallness of the light neutrino masses.

The \pSPSS builds upon the more general \SPSS \cite{Antusch:2015mia, Antusch:2016ejd}, which includes the most general \lnv extension of the \SM Lagrangian by two Majorana \HNLs $N_i$, written here as left-chiral four-component spinor fields, plus potentially other (pseudo-Dirac pairs of) Majorana \HNLs which are assumed to be negligible for collider phenomenology.
The Lagrangian density reads
\begin{equation}
    \mathcal L_{\SPSS}
    = i \widebar{N_i^c} \spd N_i^{}
    - y_{i\alpha} \widebar{N_i^c} \widetilde H^\dagger \ell_\alpha^{}
    - m_M^{} \widebar{N_1^c} N_2^{} - \mu_M^\prime \widebar{N_1^c} N_1^{} - \mu_M^{} \widebar{N_2^c}N_2^{} + \dots
    + \hc,
\end{equation}
where $y_{i\alpha}$ contains the Yukawa couplings between the sterile neutrinos, the \SM Higgs doublet~$H$ and the lepton doublets $\ell_\alpha$.
The dots capture the subleading effects of further \HNLs.
After \EWSB, the $5\times5$ part of the neutrino mass matrix, in terms of the neutrino interaction eigenstates $n = (\nu_e, \nu_\mu, \nu_\tau, N_1, N_2)^\trans$, is given by
\begin{align}
    \mathcal L_{\SPSS} &\supset - \frac12 \widebar{n^c} M_n n + \hc
    &
    M_n &= \begin{pmatrix}
        0     & m_D^\trans & \mu_D^\trans \\
        m_D^{}   & \mu_M^\prime     & m_M^{}          \\
        \mu_D^{} & m_M^{}        & \mu_M^{}
    \end{pmatrix},
\end{align}
where the masses $m_{D\alpha}^{} = v y_{1\alpha}^{}$ and $\mu_{D\alpha}^{} = vy_{2\alpha}^{}$ are generated after the Higgs boson obtains its \VEV, $v \approx \qty{174}{GeV}$.

As long as $\mu_D^{}$, $\mu_M^{}$, $\mu_M^\prime$ are much smaller than $m_M^{}$, an additional \LNLS is only mildly violated, and these small breaking parameters generate the light neutrino masses and a mass splitting $\Delta m$ between the two heavy neutrinos
\begin{equation}\label{eq:hnl-masses}
    m_{\nicefrac{4}{5}} = m_M^{} \left(1 + \frac12 \abs\theta^2\right) \pm \frac12 \Delta m,
\end{equation}
where $\theta = \flatfrac{m_D^{}}{m_M^{}}$ is the active-sterile mixing parameter.
At \LO, the \NNOs are described by the mass splitting $\Delta m$ and an additional damping parameter $\lambda$ capturing decoherence effects.
In total, to describe the collider phenomenology of a dominant pseudo-Dirac pair of \HNLs,
the required parameters are $m_M^{}$, $\Delta m$, $\theta_\alpha$, $\lambda$.
Since the average distance between two electrons within a bunch at the \FCCee is of order \unit[3]{nm} \cite{FCC:2018evy}, one can show following \cite{Antusch:2023nqd} that for the here considered \BM parameter points listed in \cref{tab:benchmark-models} the damping of \NNOs can be neglected, \ie one can set $\lambda=0$.
Hence the relevant physics can be described using the mean of the masses introduced in \cref{eq:hnl-masses}
\begin{equation}
m=\frac{m_4+m_5}{2},
\end{equation}
the mass splitting, and the active-sterile mixing angle.

\begin{table}
\begin{tabular}{cd{5.1}d{3.4}} \toprule
\BM       & \header{$\Delta m / \unit{\mu eV}$} & \header{$c\tau_\text{osc} / \unit{mm}$} \\\midrule
\ostyle 1 & 10.0                                & 124                                     \\
\ostyle 2 & 82.7                                & 15.0                                    \\
\ostyle 3 & 743                                 & 1.67                                    \\
\ostyle 4 & 42300                               & 0.0293                                  \\
\bottomrule \end{tabular}
\caption{
The \BM parameter points under study have a Majorana mass of $m_M^{} = \unit[14]{GeV}$ and active-sterile mixing parameters $(\theta_e,\theta_\mu,\theta_\tau)=(0,3,0) \times 10^{-4}$, such that $\abs{\theta}^2= \num{9\times10^{-8}}$, yielding a decay width of $\Gamma = \unit[22.6]{\mu eV}$.
}\label{tab:benchmark-models}
\end{table}

For the purpose of resolving \NNOs, the \BM points in \cref{tab:benchmark-models} will be considered, sharing the Majorana mass $m_M=\qty{14}{GeV}$, active-sterile mixings $(\theta_e,\theta_\mu,\theta_\tau) = (0,3,0) \times 10^{-4}$ such that $\abs\theta^2 = \num{9\times10^{-8}}$, leading to a decay width of $\Gamma=\qty{22.6}{\mu eV}$.
\BMs 3 and 4 correspond to the minimal linear seesaw models with a single pseudo-Dirac \HNL and inverse and normal light neutrino mass hierarchies, respectively \cite{Antusch:2022ceb}.
The mass splittings in the other two \BMs lead to larger oscillation times.
They may emerge in less minimal models with \eg multiple pseudo-Dirac pairs.

\section{Angular-dependent oscillations}\label{sec:observables}

\resetacronym{FBA}

Before we turn to the discussion of observables for pseudo-Dirac \HNLs, we summarise what is known about the angular-dependence of the \HNL production cross section from $Z$ decays at $e^+e^-$ colliders for the often considered benchmark scenarios of pure Dirac \HNLs and single Majorana \HNLs.
For the derivation of the angular-dependence, it is crucial that due to parity violation, the $Z$ bosons produced in $e^+e^-$ collisions show a polarisation of
\begin{align}
    P_Z &= - \Delta \gamma, &
    \Delta \gamma &= \gamma_L^{} - \gamma_R^{} \approx 0.1494,
\end{align}
where
\begin{align}
    \gamma_L^{} &= \frac{g_L^2}{g_L^2 + g_R^2}, &
    \gamma_R^{} &= \frac{g_R^2}{g_L^2 + g_R^2}, &
    \gamma_L^{} + \gamma_R^{} &= 1,
\end{align}
while $g_L^{} = 1 - 2\sin^2\theta_W$ and $g_R^{} = 2 \sin^2\theta_W$ are the left- and right-chiral couplings of the charged leptons to the $Z$ boson.

From the $Z$ polarisation, one obtains that the angular-dependence of the production cross section for a single heavy Majorana neutrino is given by the probability density \cite{Blondel:2021mss}
\begin{equation}\label{eq:p-majorana-angular-dependence}
    P^M(\cos\theta)
    := \frac{1}{\sigma^M} \dv[\sigma^M(\cos\theta)]{\cos\theta}
    = \frac34 \frac{m_Z^2 (1 + \cos^2\theta) + m^2\sin^2\theta}{2m_Z^2+m^2},
\end{equation}
where $\theta$ is the angle from the beam axis to the produced \HNL.

In the following, we consider the decay of a long-lived \HNL into a charged lepton or antilepton in association with an off-shell $W$ boson, which subsequently decays hadronically.
With the (anti)neutrinos escaping detection, we may consider as observable the probability density $P_{\ell^\mp}(\cos\theta)$ for the \HNL decaying into either $\ell^-$ or $\ell^+$.
Since a single Majorana \HNL decays with the same probability into a charged lepton or antilepton, we have
\begin{equation}
    P^M_{\ell^-}(\cos\theta) = P^M_{\ell^+}(\cos\theta) = \frac12 P^M(\cos\theta).
\end{equation}

For a Dirac \HNL consisting of two exactly mass degenerate Majorana neutrinos, one obtains an angular-dependence of the production cross section for a heavy neutrino $N$ and heavy antineutrino $\widebar N$ of \cite{Blondel:2021mss}
\begin{equation}\label{eq:p-dirac-angular-dependence}
    P^D_{\nicefrac{N}{\widebar N}}(\cos\theta)
    = \frac34 \frac{m_Z^2 \left[\gamma_{\nicefrac{R}{L}} (1-\cos\theta)^2 + \gamma_{\nicefrac{L}{R}} (1+\cos\theta)^2\right] + m^2\sin^2\theta}{2m_Z^2+m^2}.
\end{equation}
Although the \HNLs cannot be directly observed, their particle-antiparticle nature can be determined from the charged lepton in the final state.
By definition, a heavy neutrino $N$ decays into a charged lepton, whereas a heavy antineutrino $\widebar N$ decays into a charged antilepton.
For the observable $P_{\ell^\mp}(\cos\theta)$, this implies that
\begin{equation}
    P^D_{\ell^\mp}(\cos\theta) = P^D_{\nicefrac{N}{\widebar N}}(\cos\theta),
\end{equation}
showing a difference in the distributions for \HNLs decaying into charged leptons and antileptons, resulting from the initial $Z$ polarisation.

In the case of the Dirac \HNL, we define the normalised sum and difference of the probability densities of the neutrino-antineutrino pair
\begin{equation}\label{eq:p-pseudodirac-angular-and-time-dependence-mean-and-amplitude}
    P_N^\pm(\cos\theta) := \frac{P^D_N(\cos\theta) \pm P^D_{\widebar N}(\cos\theta)}2 ,
\end{equation}
such that
\begin{align}
    P_N^+(\cos\theta)
    &= P^M(\cos\theta)
    &
    P_N^-(\cos\theta)
    &= \frac32 \frac{m_Z^2}{2m_Z^2 + m^2} \Delta\gamma \cos\theta .
\end{align}
When these probability densities are evaluated in the forward and backward direction
\begin{align} \label{eq:FBA}
S^\text{FB} &:= P_N^+(\pm1)
 = \frac32 \frac{m_Z^2}{2m_Z^2 + m^2}, &
A_{\nicefrac{N}{\widebar{N}}}^\text{FB} &:= P_N^-(\pm1) = \pm S^\text{FB} \Delta \gamma,
\end{align}
they generate a direction-independent symmetric expression and a \fba term \cite{Blondel:2021mss}.

However, the single Majorana \HNL and the pure Dirac \HNL discussed above are not realistic \BMs for neutrino mass generation.
A pure Dirac \HNL leaves the light neutrinos exactly massless, while a single Majorana neutrino cannot explain the observed mass squared differences.
Furthermore, taking into account the constraints on the one light neutrino mass it generates implies that its Yukawa couplings (or active-sterile mixing angles) are too small for tests at colliders.
It is, therefore, interesting to study the case of pseudo-Dirac \HNLs, which, as discussed above, are a generic feature of collider-testable low-scale seesaw models capable of explaining the observed light neutrino masses.

\subsection{Angular- and time-dependent distributions} \label{sec:observables-full-dependence}

The angular probabilities for pseudo-Dirac \HNLs decaying into charged leptons $\ell^\mp$ can be obtained from the ones for a Dirac \HNL by exploiting that the produced heavy neutrino $N$ and heavy antineutrino $\widebar N$ oscillate as a function of the \HNL's proper time $\tau$ after production.
The \LO formulae for the oscillation probabilities can be written as \cite{Antusch:2022ceb}
\begin{equation}
    P_\text{osc}^{\nicefrac{\LNC}{\LNV}}(\tau) = \frac{1 \pm \cos(\Delta m \tau)\exp(-\lambda)}{2},
\end{equation}
where \LNC stands for the survival probability, \ie the probability that
the neutrino's (anti)particle nature remains unchanged, whereas \LNV stands for the oscillation probability, \ie the probability that a $N$ ($\widebar N$) oscillates into a $\widebar N$ ($N$), violating lepton number.
Furthermore, the probability density for the decay of the \HNL at proper time $\tau$ as a function of the decay width $\Gamma = \Gamma(m,\theta)$ is given by
\begin{equation}
    P_\text{decay}(\tau) = - \dv \tau \exp\left(- \Gamma \tau\right) = \Gamma \exp\left(- \Gamma \tau\right).
\end{equation}
Combining them yields the probability density
\begin{equation}\label{eq:probability-displacement}
    P_{\nu\ell}^{\nicefrac{\LNC}{\LNV}}(\tau) = P^{\nicefrac{\LNC}{\LNV}}_\text{osc}(\tau) P_\text{decay}(\tau),
\end{equation}
for the survival/oscillation of the unstable $N$ or $\widebar N$ until it decays at proper time $\tau$.

\begin{figure}
\begin{panels}{4}
 \includetikz{BM1-Illustration}
 \caption{\BM{}1}
 \panel
 \includetikz{BM2-Illustration}
 \caption{\BM{}2}
 \panel
 \includetikz{BM3-Illustration}
 \caption{\BM{}3}
 \panel
 \includetikz{BM-Majorana-Illustration}
 \caption{\BM{}4}
\end{panels}
\caption{
Illustration of the probability density ratio $R_\ell(\tau, \cos \theta)$ given in \eqref{eq:rl-pseudodirac-angular-and-time-dependence} for the \BM points in \cref{tab:benchmark-models}.
The angle $\theta = 0$ corresponds to the direction along the $z$ axis, \ie the direction of the electron momentum.
The radius corresponds to a proper time within the range $c\tau \in \unit[{{[0,15]}}]{mm}$.
Blue colour indicates $R_\ell(\tau, \cos \theta) > 1$, \ie that the \HNLs decay preferably into leptons, whereas red colour indicates $R_\ell(\tau, \cos \theta) < 1 $, \ie that the \HNLs decay preferably into antileptons.
For \BM{}4, there are about $500$ oscillations between red and blue for the shown range in $c\tau$.
Considering the reconstruction resolution, they are replaced by plain violet, indicating an averaged probability density ratio of one.
} \label{fig:illustration}
\end{figure}

One can now observe decays of a produced \HNL into \eg $\ell^-$ either by producing a $N$ that survives or by producing a $\widebar N$ that has oscillated into a $N$, and analogously for $\ell^+$ final states.
This yields decaying oscillations around $P_N^+(\cos\theta)$ with an amplitude $P_N^-(\cos\theta)$, which are defined in \eqref{eq:p-pseudodirac-angular-and-time-dependence-mean-and-amplitude}
\begin{equation}\label{eq:p-pseudodirac-angular-and-time-dependence}\begin{split}
    P_{\ell^\mp}(\tau, \cos\theta) &
    = P_{\nu\ell}^{\LNC}(\tau) P_{\nicefrac{N}{\widebar N}}(\cos\theta) + P_{\nu\ell}^{\LNV}(\tau) P_{\nicefrac{\widebar N}{N}}(\cos\theta) \\&
    = P_\text{decay}(\tau) \left[ P_N^+(\cos\theta) \pm P_N^-(\cos\theta) \Delta P_\text{osc}(\tau) \right],
\end{split}\end{equation}
where
\begin{equation}
    \Delta P_\text{osc}(\tau)
    := P_\text{osc}^\text{LNC}(\tau) - P_\text{osc}^\text{LNV}(\tau)
    = \cos(\Delta m \tau)\exp(-\lambda).
\end{equation}
This method can also be applied to other observables known for Dirac \HNLs.
Here we will focus on $P_{\ell^\mp}(\tau, \cos\theta)$ and observables that can be derived from it.
For instance, one can define the ratio
\begin{equation}\label{eq:rl-pseudodirac-angular-and-time-dependence}
    R_\ell(\tau, \cos \theta)
    = \frac{P_{\ell^-}(\tau, \cos\theta)}{P_{\ell^+}(\tau, \cos\theta)}
    = 1 + 2\frac{P_N^-(\cos\theta)}{P_N^+(\cos\theta) \Delta P_\text{osc}^{-1}(\tau) - P_N^-(\cos\theta)}
\end{equation}
\ie the angular- and time-dependent probability ratio density for the \HNLs to decay into a charged lepton or antilepton.
This probability ratio is illustrated in \cref{fig:illustration} for the \BM points defined in \cref{tab:benchmark-models}.

\subsection{Time-integrated distributions} \label{sec:observables-time-integrated}

Even when the $\tau$ oscillations of the observables defined in \eqref{eq:p-pseudodirac-angular-and-time-dependence,eq:rl-pseudodirac-angular-and-time-dependence} cannot be resolved experimentally, they can nevertheless have significant effects.
Integrating the former from zero to infinity over the proper time $\tau$ results in
\begin{equation}\label{eq:p-time-integrated}
    P_{\ell^\mp}(\cos\theta)
    := \int_0^\infty P_{\ell^\mp}(\tau, \cos\theta) \d\tau
    = P_N^+(\cos\theta) \pm P_N^-(\cos\theta) \Delta P_{\nu\ell},
\end{equation}
and the resulting ratio between events having a lepton or an antilepton in its final state is
\begin{equation}\label{eq:r-time-integrated}
    R_\ell(\cos \theta)
    := \frac{P_{\ell^-}(\cos\theta)}{P_{\ell^+}(\cos\theta)}
    = 1 + 2 \frac{P_N^-(\cos\theta)}{P_N^+(\cos\theta) \Delta P_{\nu\ell}^{-1} -  P_N^-(\cos\theta)},
\end{equation}
where the total difference between the time-integrated probabilities for \lnc and \lnv events
\begin{equation}
    \Delta P_{\nu\ell}
    := \int_0^\infty P_{\nu\ell}^\text{LNC}(\tau) - P_{\nu\ell}^\text{LNV}(\tau) \d\tau
    = \frac{\Gamma^2}{\Gamma^2 + \Delta m^2} e^{-\lambda}
\end{equation}
encapsulates the interplay of the oscillations governed by the mass splitting $\Delta m$ and decay at rate $\Gamma$.
It may allow $\Delta m$ to be extracted even if oscillations cannot be resolved.
\footnote{
  We remark that, in order to take into account the detector geometry, the integration boundaries should be adapted, as discussed in \cite{Antusch:2022ceb}.
}

Finally, we note that the time-independent probability density of the pure Dirac case in \cref{eq:p-dirac-angular-dependence}, where lepton number is conserved, can be recovered when the mass splitting is taken to zero
\footnote{
    Since damping describes an effect intimately related to oscillations, it vanishes whenever they are neglected under the limits of vanishing and infinite mass splitting.
}
\begin{equation}
    P_{\ell^\mp}(\cos\theta)
    \xrightarrow[\lambda \to 0]{\Delta m \to 0} P_N^+(\cos\theta) \pm P_N^-(\cos\theta)
    = P^D_{\nicefrac{N}{\widebar N}}(\cos\theta) .
\end{equation}
Likewise, the angular-dependence for a single Majorana \HNL given in \cref{eq:p-majorana-angular-dependence} is recovered for the limit in which $\Delta m$ goes to infinity
\begin{equation}
    P_{\ell^\mp}(\cos\theta)
    \xrightarrow{\Delta m \to \infty} P_N^+(\cos\theta)
    = P^M(\cos\theta).
\end{equation}
Pseudo-Dirac \HNLs feature angular dependencies between these two limiting cases.

\subsection{Angular-integrated distributions} \label{sec:observables-angular-integrated}

The observables discussed so far are densities w.r.t.\ the angle $\theta$.
In a realistic experimental situation, intervals of $\theta$ are considered, \ie one defines angular-integrated variables.
From \eg \eqref{eq:p-pseudodirac-angular-and-time-dependence} and \eqref{eq:rl-pseudodirac-angular-and-time-dependence} one arrives at
\begin{equation}\label{eq:p-arbitrary-angular-integrated}\begin{split}
    P_{\ell^\mp}^{[\theta_\text{min},\theta_\text{max}]}(\tau) &
    := \int_{\cos\theta_\text{min}}^{\cos\theta_\text{max}} P_{\ell^\mp}(\tau,\cos\theta) \d \cos \theta
    \\&
    = P_\text{decay}(\tau) \left[ P_N^{+[\theta_\text{min},\theta_\text{max}]} \pm P_N^{-[\theta_\text{min},\theta_\text{max}]} \Delta P_\text{osc}(\tau) \right],
\end{split}\end{equation}
where
\begin{align}\label{eq:p-arbitrary-angular-integrated-mean-and-amplitude}
    P_N^{+[\theta_\text{min},\theta_\text{max}]}
    &:= 3 x_- + 2 S^\text{FB} (x_+  - 2 x_-),
    &
    P_N^{-[\theta_\text{min},\theta_\text{max}]}
    &:= A^\text{FB}_N \frac{\cos^2\theta_\text{max} - \cos^2\theta_\text{min}}2
\end{align}
and
\begin{equation}\label{eq:p-arbitrary-angular-integrated-mean-and-amplitude-x}
    x_\pm
    := \pm \frac{\cos\theta_\text{max} - \cos\theta_\text{min}}2 \frac{\cos^2\theta_\text{max} + \cos\theta_\text{max}\cos\theta_\text{min} + \cos^2\theta_\text{min} \pm 3}6.
\end{equation}
The ratio between events with leptons and antileptons therefore reads
\begin{equation}\label{eq:r-arbitrary-angular-integrated}
    R_\ell^{[\theta_\text{min},\theta_\text{max}]}(\tau)
    :=  \frac{P_{\ell^-}^{[\theta_\text{min},\theta_\text{max}]}(\tau)}{P_{\ell^+}^{[\theta_\text{min},\theta_\text{max}]}(\tau)}
    = 1 + 2 \frac{P_N^{-[\theta_\text{min},\theta_\text{max}]}}{P_N^{+[\theta_\text{min},\theta_\text{max}]} \Delta P_\text{osc}^{-1}(\tau) - P_N^{-[\theta_\text{min},\theta_\text{max}]}} .
\end{equation}

Interesting special cases of \cref{eq:p-arbitrary-angular-integrated,eq:r-arbitrary-angular-integrated} arise when the integration boundaries are chosen to specify the positive and negative $z$ directions, corresponding to
\begin{equation}\label{eq:p-angular-integrated}
    P_{\ell^\mp}^{[\nicefrac{\pi}{2},0]}(\tau)
    = P_{\ell^\pm}^{[\pi,\nicefrac{\pi}{2}]}(\tau)
    = \frac{1 + A_{\nicefrac{N}{\widebar{N}}}^\text{FB} \Delta P_\text{osc}(\tau)}2 P_\text{decay}(\tau)
\end{equation}
and
\begin{align}\label{eq:r-angular-integrated}
    R_\ell^{[\nicefrac{\pi}{2},0]}(\tau)
    &= \frac{P^{[\nicefrac{\pi}{2},0]}_{\ell^-}(\tau)}{P^{[\nicefrac{\pi}{2},0]}_{\ell^+}(\tau)}
    = \frac{1 + A_{\ell^-}^\text{FB}(\tau)}{1 + A_{\ell^+}^\text{FB}(\tau)},
    &
    R_\ell^{[\pi,\nicefrac{\pi}{2}]}(\tau)
    &= \frac{P^{[\pi,\nicefrac{\pi}{2}]}_{\ell^-}(\tau)}{P^{[\pi,\nicefrac{\pi}{2}]}_{\ell^+}(\tau)}
    = \frac{1 - A_{\ell^-}^\text{FB}(\tau)}{1 - A_{\ell^+}^\text{FB}(\tau)},
\end{align}
where the time-dependent \FBAs are
\begin{equation}\label{eq:fba-time-dependent}
    A_{\ell^\mp}^\text{FB} (\tau)
    := \frac{P_{\ell^\mp}^{[\nicefrac{\pi}{2},0]}(\tau) - P_{\ell^\mp}^{[\pi,\nicefrac{\pi}{2}]}(\tau)}{P_{\ell^\mp}^{[\nicefrac{\pi}{2},0]}(\tau) + P_{\ell^\mp}^{[\pi,\nicefrac{\pi}{2}]}(\tau)}
    = A_{\nicefrac{N}{\widebar{N}}}^\text{FB} \Delta P_\text{osc}(\tau).
\end{equation}

In the absence of \NNOs at $\tau=0$, the \FBA for pure heavy Dirac neutrinos and antineutrinos \eqref{eq:FBA} are recovered with charged leptons and antileptons, respectively
\begin{equation}\label{eq:fba-time-dependent-limits}
    A_{\ell^\mp}^\text{FB}(\tau=0)
    = A_{\nicefrac{N}{\widebar{N}}}^\text{FB} .
\end{equation}
However, if heavy neutrino decays after a finite lifetime are considered, the \NNOs induce oscillations of $A_{\ell^-}^\text{FB} (\tau)$ and $A_{\ell^+}^\text{FB} (\tau)$ as a function of $\tau$, with the same oscillation time as the \NNOs themselves.

Although the aforementioned oscillations can and will be simulations at the reconstructed level in \cref{sec:simulations}, statistical noise can be reduced by computing their difference as
\begin{equation}\label{eq:fba-time-dependent-combined}
    A_\ell^\text{FB} (\tau)
    := \frac{A_{\ell^-}^\text{FB} (\tau) - A_{\ell^+}^\text{FB} (\tau)}{2}
    = A_N^\text{FB} \Delta P_\text{osc}(\tau).
\end{equation}

\subsection{Angular- and time-integrated distributions}\label{sec:observables-angular-time-integrated}

Finally, integrating the probability density \eqref{eq:p-pseudodirac-angular-and-time-dependence} in both angle and proper time yields
\begin{equation}\label{eq:p-arbitrary-angular-and-time-integrated}
    P_{\ell^\mp}^{[\theta_\text{min},\theta_\text{max}]}
    := \int_0^\infty \int_{\cos\theta_\text{min}}^{\cos\theta_\text{max}} P_{\ell^\mp}(\tau,\cos\theta) \d \cos \theta \d\tau
    = P_N^{+[\theta_\text{min},\theta_\text{max}]} \pm P_N^{-[\theta_\text{min},\theta_\text{max}]} \Delta P_{\nu\ell}
\end{equation}
and the ratio between events with charged leptons and antileptons becomes
\begin{equation}\label{eq:r-arbitrary-angular-and-time-integrated}
    R_\ell^{[\theta_\text{min},\theta_\text{max}]}
    :=  \frac{P_{\ell^-}^{[\theta_\text{min},\theta_\text{max}]}}{P_{\ell^+}^{[\theta_\text{min},\theta_\text{max}]}}
    = 1 + 2 \frac{P_N^{-[\theta_\text{min},\theta_\text{max}]}}{P_N^{+[\theta_\text{min},\theta_\text{max}]} \Delta P_{\nu\ell}^{-1} - P_N^{-[\theta_\text{min},\theta_\text{max}]}}.
\end{equation}

Taking once more the $\theta$ intervals corresponding to the positive and negative $z$ directions yields the probabilities
\begin{equation}\label{eq:p-angular-and-time-integrated}
    P^{[\nicefrac{\pi}{2},0]}_{\ell^\mp}
    = P^{[\pi,\nicefrac{\pi}{2}]}_{\ell^\pm}
    = \frac{1 + A_{\nicefrac{N}{\widebar{N}}}^\text{FB} \Delta P_{\nu\ell}}2,
\end{equation}
which leads to the probability ratios
\begin{align}\label{eq:r-angular-and-time-integrated}
    R_\ell^{[\nicefrac{\pi}{2},0]}
    &= \frac{P^{[\nicefrac{\pi}{2},0]}_{\ell^-}}{P^{[\nicefrac{\pi}{2},0]}_{\ell^+}}
    = \frac{1 + A_{\ell^-}^\text{FB}}{1 + A_{\ell^+}^\text{FB}},
    &
    R_\ell^{[\pi,\nicefrac{\pi}{2}]}
    &= \frac{P^{[\pi,\nicefrac{\pi}{2}]}_{\ell^-}}{P^{[\pi,\nicefrac{\pi}{2}]}_{\ell^+}}
    = \frac{1 - A_{\ell^-}^\text{FB}}{1 - A_{\ell^+}^\text{FB}},
\end{align}
where the time-integrated \FBAs are
\begin{equation}\label{eq:fba-time-integrated}
    A_{\ell^\mp}^\text{FB}
    = A_{\nicefrac{N}{\widebar{N}}}^\text{FB} \Delta P_{\nu\ell}.
\end{equation}
Sending the mass splitting to zero and infinity recovers the known \FBAs for the pure Dirac \HNL given in \eqref{eq:FBA} and the single Majorana \HNL
\begin{align}
    A_{\ell^\mp}^\text{FB} &\xrightarrow[\lambda \to 0]{\Delta m \to 0} A_{\nicefrac{N}{\widebar{N}}}^\text{FB}, &
    A_{\ell^\mp}^\text{FB} &\xrightarrow{\Delta m \to \infty} 0.
\end{align}

\section{Simulation}\label{sec:simulations}

\begin{figure}
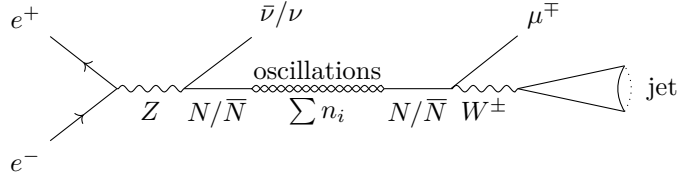

\includetikz*{Z-Pole-Diagram}
\caption{
Diagram depicting the production, oscillation, and decay of an \HNL.
From the $Z$ boson, the heavy neutrino (antineutrino) interaction eigenstate $N$ ($\widebar N$) is produced in association with a light antineutrino (neutrino).
$N$ and $\widebar N$ are linear combinations of the heavy mass eigenstates $n_4$ and $n_5$.
The interference of the propagating mass eigenstates induces oscillations between $N$ and $\widebar N$ as a function of the \HNL lifetime.
Due to the oscillations, the \HNL can decay into a (displaced) muon or antimuon (from the decay of a $N$ or $\widebar N$, respectively).
We indicate that the process is initiated by electron-positron collisions and, for our parameter points, the two final quarks, originating from the hadronic $W$ decay, immediately hadronise into a single jet.
Due to the \NNOs, the total process can be \lnc or \lnv.
}\label{fig:z-pole-diagram}
\end{figure}

With the goal of simulating the process in \cref{fig:z-pole-diagram}, the \pSPSS 1.0 \feynrules{} implementation was adapted to account for the definite lepton number with which light neutrinos are produced by modelling them as Dirac-type \cite{FR:pSPSS_Dirac}.
The tiny Majorana masses they receive in the (p)\SPSS from the small breaking of the \LNLS can be neglected in the process under consideration.
\feynrulesver{} is then used to parse the modified \pSPSS model file into a UFO folder that \madgraphver{} can use to generate parton level events for the process in \cref{fig:z-pole-diagram}:
\begin{verbatim}
generate e- e+ > Z > vmu nn, (nn > mu j j)
\end{verbatim}

Since the heavy neutrinos are an explicit intermediate state, the correct interference between the mass eigenstates can be introduced by the \madgraph{} patch published in \cite{Antusch:2022ceb} with the inclusion of light neutrinos for the calculation of the particle-antiparticle derivation.
Subsequently, \pythiaver{} and \delphesver{} with a modified \IDEA card are called to hadronise, shower, and simulate detector effects on the generated events.

The expected number of $Z\to N\nu$ decays during the $Z$ pole run of the \FCCee is the product of the process' cross section by the run's luminosity.
Retrieving the former from \madgraph{} and estimating the latter from the latest predictions for the total number of events \cite{PhysicsPerformanceMeeting2023}, around \num{29700} heavy neutrinos should be produced.

As the heavy neutrinos in the \BMs under study are long-lived, the muon and two quarks are expected to be produced from a secondary vertex displaced from the primary vertex of the $e^-e^+$ interaction.
The final state is then composed of an undetectable light neutrino, a single displaced muon, and hadronic activity from the two quarks.
Hence, only events with a single jet and a displaced (anti)muon are considered.
For this purpose, a particle is defined as displaced if its origin is at least \qty{400}{\mu m} away from the primary vertex \cite{Drewes:2022rsk} while remaining inside a cylinder with half the tracker size in each direction, corresponding to \qty{1}{m} longitudinally and transversely for the \IDEA detector.
Events surviving these cuts allow for the neutrino's momentum and proper lifetime to be reconstructed from its decay products and the position of the secondary vertex.

As previously mentioned, it was necessary to modify the default \IDEA card provided by \delphes.
Indeed, a bug was identified causing muons to be simultaneously detected as such and also included into jets.
To address this problem, the \code{EFlowFilter} responsible for removing electrons and muons from the energy flow was copied over from the default CMS card and provided as an input to \code{PhotonIsolation}, \code{ElectronIsolation}, \code{MuonIsolation}, and \code{FastJetFinder}.
Additionally, the muon isolation $\Delta R$ cut in the card was reduced from $\Delta R_\text{max}=0.5$ to $\Delta R_\text{max}=0.1$.

\section{Results}\label{sec:results}

\begin{figure}
\includetikz{groupplot}
    \begin{panels}{.1}
        \strut
        \panel{.225}
        \caption{\BM{}1} \label{column:1}
        \panel{.225}
        \caption{\BM{}2} \label{column:2}
        \panel{.225}
        \caption{\BM{}3} \label{column:3}
        \panel{.225}
        \caption{\BM{}4} \label{column:4}
    \end{panels}
\caption{
    Comparison between theory predictions (\ref*{plot:muon-func}, \ref*{plot:amuon-func}, and \ref*{plot:combined-func}) and simulated results (\ref*{plot:muon-data}, \ref*{plot:amuon-data}, and \ref*{plot:combined-data}) for the observables $P_{\mu^\mp}(\cos\theta)$, $A_{\mu^\mp}^\text{FB} (\tau)$, and $A_\mu^\text{FB} (\tau)$ define in \cref{eq:p-time-integrated,eq:fba-time-dependent,eq:fba-time-dependent-combined}, respectively.
    The four \BM points defined in \cref{tab:benchmark-models} are shown in columns \subref{column:1} to \subref{column:4}.
    The first two rows show the data available from $c\tau= \unit[0\text{ to }15]{mm}$ with solid blue lines and downward-pointing blue marks (\ref*{plot:muon-func} and \ref*{plot:muon-data}) corresponding to $\mu^-$-observables and red dashed lines and upward-pointing red marks (\ref*{plot:amuon-func} and \ref*{plot:amuon-data}) to $\mu^+$-observables.
    The third row shows the combined decay asymmetry $A_\mu^\text{FB} (\tau)$ for the first complete oscillation available for which there is data.
} \label{fig:results}
\end{figure}

The results of our simulations are presented in \cref{fig:results}.
In the two upper rows of plots, solid blue lines and downward-pointing blue marks (dashed red lines and upward-pointing red marks) are theory predictions and reconstructed observables for events with muons (antimuons), respectively.
The upper row shows that the time-integrated distributions $P_{\ell^\mp}(\cos\theta)$ of our simulations of reconstructed heavy (anti)neutrinos over $\cos\theta$ (with $40$ bins) reproduce well the analytical prediction  \eqref{eq:p-time-integrated}.
For \BM{}1, $P_{\ell^\mp}(\cos\theta)$ features a significant asymmetry between the distributions for muons and antimuons, resembling those for a pure Dirac \HNL \cite{Blondel:2021mss}.
While this difference is still noticeable in the theory prediction for \BM{}2, it is no longer visible in the remaining \BMs, which thus resemble the distribution for a single Majorana \HNL.

In the second row of \cref{fig:results}, the simulated time-dependent \FBA $A_{\ell^\mp}^\text{FB} (\tau)$ with $40$ bins in $c\tau\in \unit[{{[0,15]}}]{mm}$ is compared to the theory prediction \eqref{eq:fba-time-dependent}.
From \BM{}1 to \BM{}4, the mass splitting increases, and the oscillation period decreases correspondingly.
While \BM{}1 generates an almost flat distribution over the chosen $c\tau$ range, \BM{}2 shows significant oscillations that are potentially observable.
Resolving the oscillations with the chosen resolution becomes challenging for \BM{}3 and nearly impossible for \BM{}4.

In order to make this analysis more pronounced, the third row shows the theory predictions and the reconstructed results from our simulations for the combined decay asymmetry \eqref{eq:fba-time-dependent-combined}.
For \BM{}1, the \HNLs decay while the first oscillation is just starting, but,  from the reconstructed data, one may nevertheless find evidence for a decrease of $A_\mu^\text{FB} (\tau)$ with $\tau$.
\BM{}4 is an example where the oscillations of $A_\mu^\text{FB} (\tau)$ seem not resolvable.
The oscillation length $c\tau_\text{osc}$ is so small that each bin has very few events, leading to a large spread of the reconstructed data points, some of which lie outside the chosen $y$-axis range.
For \BM{}2, the reconstructed data clearly shows the oscillatory behaviour of the theory prediction, providing examples with good prospects to discover the oscillations at the \FCCee and to measure the heavy mass splitting $\Delta m$.
This oscillatory pattern is also visible for \BM{}3, though less clearly than for \BM{}2.

\section{Conclusion}\label{sec:conclusions}

This paper discussed the impact of \NNOs on \HNL searches at proposed electron-positron colliders such as the \FCCee.
We have considered collider-testable low-scale seesaw scenarios, where the \HNLs form pseudo-Dirac pairs.
\LNV in such scenarios can be understood in terms of \NNOs.
The process we have focused on is the production of \HNLs from $Z$ decays along with a light neutrino or antineutrino that escapes detection and their decay into a charged lepton or antilepton and an off-shell $W$ boson.
We have discussed how the oscillations modify the angular distribution of the reconstructed heavy (anti)neutrinos, which, for pseudo-Dirac \HNLs, develops an oscillatory signature as a function of the \HNL lifetime.
We have also discussed the oscillations of angular-integrated observables such as the time-dependent \FBAs, as well as time- and angular-integrated ones, \eg $R_\ell^{[\theta_\text{min},\theta_\text{max}]}$, the ratio of the total number of \HNL decays into $\ell^-$ over the ones into $\ell^+$, in an interval of the angle $\theta$ between the \HNL and the beam axis.

For collider simulations of \HNLs from $Z$ boson decays at the \FCCee, we have used the \pSPSS benchmark scenario \cite{Antusch:2022ceb} and an adapted version of the provided \software{MadGraph} patch.
For four selected \BM points, we have simulated the oscillating \FBAs and the time-integrated distribution of reconstructed heavy (anti)neutrinos over $\theta$.
We have shown that for \BM{}2, and possibly \BM{}3, there are good prospects to resolve the oscillation of the combined \FBA $A_{\mu}^\mathrm{FB} (\tau)$, which would allow to measure the heavy mass splitting $\Delta m$.
For three of the \BM points (\BM{}2, \BM{}3, \BM{}4), where the oscillation time is much shorter than the lifetime, the time-integrated distributions show no significant asymmetries, as one would also expect for a single Majorana \HNL.
One of the benchmark points, \BM{}1, which features a larger oscillation time, has a notably asymmetric distribution, as one would also expect for a pure Dirac \HNL.
The characteristics of the two previously studied \emph{single Majorana \HNL} and \emph{Dirac \HNL} \BM scenarios are thus recovered as limiting cases of the pseudo-Dirac scenario.

After the discovery of \HNLs, the question will arise whether they violate lepton number.
At $pp$ colliders, this can be addressed straightforwardly by counting the total lepton number of the final states, ensuring no lepton number escapes detection as light neutrinos.
At the \FCCee, this is challenging without assuming a particular theory model.
For example, even when an observable allows to distinguish between the two often used \BM scenarios with a pure Dirac \HNL (being \lnc) or one single Majorana \HNL (being \lnv), this does not allow a model-independent statement about the violation or conservation of lepton number.
On the contrary, observing a time-dependent lepton number of a decaying particle (or set of particles) would provide a clear model-independent signal of \LNV.
This observable is given \eg by $R_\ell^{[\theta_\text{min},\theta_\text{max}]}(\tau)$, for some properly chosen interval of $\theta$.
For the example of pseudo-Dirac \HNLs, it oscillates due to \NNOs.
\footnote{
  To feature oscillations, the interval should not be symmetric around $\theta = \pi/2$.
  Additionally, for $R_\ell$ to be the ratio of \HNL decays into final states with $L=+1$ over those with $L=-1$, there must be no lepton number $L$ in final states apart from the one carried by $\ell^\mp$, which requires \eg choosing a process without additional neutrinos and vetoing against $\tau$ jets.
}
However, even without any theory interpretation, the time-dependence means that no fixed lepton number can be assigned to the decaying particle(s), \ie that lepton number must be violated.

In summary, collider-testable low-scale seesaw models featuring pseudo-Dirac pairs of \HNLs, capable of explaining the observed light neutrino masses, have a rich phenomenology.
Testable \LNV is induced via \NNOs, manifesting in various observables.
Depending on the \HNL mass splitting $\Delta m$, these oscillations could be resolvable, allowing for a deeper insight into the mechanism of neutrino mass generation and the clear discovery of \LNV at the \FCCee.

\printbibliography

\end{document}